\documentclass[12pt,amssymb,amsmath,english,aps,preprint,showpacs]{revtex4}
\usepackage{hyperref}
\usepackage{subfigure}
\usepackage{graphicx}

\newcommand{\unit}[1]{\,{\rm #1}}

\newcommand{\s}{\unit{s}}

\newcommand{\Rm}{Re_{\rm m}}

\providecommand{\boldsymbol}[1]{\mbox{\boldmath $#1$}}

\usepackage{babel}

\begin{document}

\preprint{APS/123-DPP}

\title{Magnetized Ekman Layer and Stewartson Layer in a Magnetized Taylor-Couette Flow}

\author{Wei Liu\footnote{Email: wliu@lanl.gov}}

\affiliation{Center for Magnetic Self-Organization in Laboratory and Astrophysical Plasma, Los Alamos National Laboratory, Los Alamos, NM, USA 87545}

\date{\today}

\begin{abstract}
 In this paper we present axisymmetric nonlinear simulations of magnetized Ekman and Stewartson layers in a magnetized Taylor-Couette flow with a centrifugally stable angular-momemtum profile and with a magnetic Reynolds number below the threshold of magnetorotational instability. The magnetic field is found to inhibit the Ekman suction. The width of the Ekman layer is reduced with increased magnetic field normal to the end plate. A uniformly-rotating region forms near the outer cylinder. A strong magnetic field leads to a steady Stewartson layer emanating from the junction between differentially rotating rings at the endcaps. The Stewartson layer becomes thinner with larger Reynolds number and penetrates deeper into the bulk flow with stronger magnetic field and larger Reynolds number. However, at Reynolds number larger than a critical value $\sim 600$, axisymmetric,  and perhaps also nonaxisymmetric, instabilities occur and result in a less prominent Stewartson layer that extends less far from the boundary.
\end{abstract}


\maketitle

\section{Introduction}
The history of Taylor-Couette flow dates back to the 19th century. To measure viscosity, \citet{ct1890} studied flows between rotating concentric cylinders. Rayleigh's stability criterion was introduced in 1916 during his study of cyclones. \citet{tg23} extended it by including viscosity, and made quantitative predictions of instability in Couette flow.
If the cylinders were infinitely
long, the steady-state laminar solution would be the ideal
Taylor-Couette state:
\begin{equation}
\label{couette}
\Omega_0(r)=a+\frac{b}{r^{2}}\, ,
\end{equation}
in which $a=(\Omega_{2}r_{2}^{2}-\Omega_{1}r_{1}^{2})/(r_{2}^{2}-r_{1}^{2})$
and $b=r_{1}^{2}r_{2}^{2}(\Omega_{1}-\Omega_{2})/(r_{2}^{2}-r_{1}^{2})$, where $r_1$ and $r_2$ are the radius of the inner and outer cylinder, and $\Omega_1$ and $\Omega_2$ are the angular velocity of the inner and outer cylinder respectively.
Rayleigh's stability criterion states that in the unmagnetized and inviscid limit, such a flow is
linearly axisymmetrically stable if and only if the specific
angular momentum increases outwards: $ab>0$.

The study of magnetized Taylor-Couette flow began much later. \citet{ve59} and \citet{chan60} discovered that a vertical magnetic field may destabilize the flow,
provided that the angular \emph{velocity} decreases outward,
$\Omega_{2}^2<\Omega_{1}^2$, which today is called magnetorotational instability (MRI). In ideal magnetohydrodynamics (MHD), the instability takes place with an arbitrarily weak field \citep{bh91,bh98}. Experiments on magnetized Couette flow aiming to observe MRI have been performed \cite{sisan04,sgg06}, but MRI has never been conclusively demonstrated in the laboratory. Some other experiments have been proposed or are still under construction \cite{npc02,vils06,jgk01,gj02}.
The experimental geometry planned by most groups is a magnetized
Taylor-Couette flow: an incompressible liquid metal with density $\rho$, kinematic viscosity $\nu$ and magnetic resistivity $\eta$ confined between
concentric rotating cylinders, with an imposed axial and/or toroidal background magnetic
field sustained by currents external to the fluid. 

The challenge for experimentation is that liquid-metal flows
are very far from ideal on laboratory scales.  While the fluid
Reynolds number $Re\equiv \Omega_{1}r_{1}(r_{2}-r_{1})/\nu$ can be
large, the corresponding \emph{magnetic} Reynolds number
$\Rm\equiv\Omega_{1}r_{1}(r_{2}-r_{1})/\eta$ is modest or small,
because the magnetic Prandtl number $Pr_{\rm m}\equiv\nu/\eta\sim
10^{-5}-10^{-6}$ in liquid metals.  Standard MRI modes will not grow unless both
the rotation period and the Alfv\'en crossing time are shorter than
the timescale for magnetic diffusion.  This requires both $\Rm\gtrsim
1$ and $S\gtrsim 1$, where $S\equiv V_{A}(r_{2}-r_{1})/\eta$ is the
Lundquist number, in which $V_{A}=B_{z0}/\sqrt{\mu_0\rho}$ is the Alfv\'en speed and $B_{z0}$ is the imposed axial magnetic field.
Therefore, $Re\gtrsim 10^6$ and fields of several kilogauss must 
typically be achieved. Hollerbach and collaborators have discovered that MRI-like
modes may grow at much reduced $\Rm$ and $S$ in the presence of a
helical background field, a current-free combination of axial and
toroidal field \citep{hr05,rhss05}. Though \citet{sgg06} have claimed to observe this helical MRI (HMRI) experimentally, we explained the experimentally measured wave patterns to be transiently amplified disturbances launched by viscous boundary layers rather than globally unstable modes \citep{lgj07}. We also questioned the relevance of this helical MRI to astrophysics by showing that this new mode is stable for a  Keplerian rotation profile by WKB analysis in a narrow-gap geometry (see Sec.II.A of \citet{lghj06}) and by linear calculations in a wide-gap geometry (see Sec.II.B of \citet{lghj06}).  Recently \citet{rh07,pgg07} have reported that this new mode is unstable in the inductionless limit with for some boundary conditions. Under the parameters used in the \citet{rh07,pgg07} ($Rm_{\rm m}=S=Pr_{\rm m}=0$, but with finite $Re$ and Hartmann number $Ha=V_{A}(r_2-r_1)/\sqrt{\eta\nu}$), the authors are indeed taking the diffusivity to infinity $\eta\to\infty$. Note however that the combination $Ha^2/2Re = (V_A)^2/(2\Omega\eta)$, which is the Elsasser number $\Lambda$ \cite{gp71}, is also finite. The authors consider $Ha$ and $Re$ to be constant as $Pr_{\rm m}\to0$; thus if we think of $\Omega$ and $\nu$ as fixed, then the Alfven speed must scale like $\sqrt{\eta}$ as $\eta\to\infty$. So, the authors are considering a limit in which the Alfven speed is infinitely larger than the rotation speed but poorly coupled to the flow, whereas \citet{lghj06} were thinking of the resistive limit as one in which the field strength and rotation speed were held fixed as the diffusivity became infinite. The former limit is unlikely to be important in astrophysics. However, it might be achieved in a low-plasma-beta but highly resistive (weakly ionized) plasma.

In view of the large Reynolds number, the Taylor-Proudman theorem suggests that the end plates should dominate the entire flow unless a very long cylinder were used ($h/(r_2-r_1)\gg10^3$, where $h$ is the height of the cylinders.) \cite{hf04}. The no-slip boundary condition on the end plates causes an imbalance between centrifugal and pressure forces and drives Ekman circulation.  If the endcaps rotate rigidly with the outer cylinder, this circulation takes the form of inward flows along the endcaps, which turn vertically along the inner cylinder, converge at the midplane, and depart the cylinder in a radial jet \cite{kjg04}. This Ekman circulation, and especially the jet, transport angular momentum efficiently and reduce the free energy available for shear-driven instabilities \cite{kjg04}. Both effects are unfavorable for laboratory demonstration of MRI. The Princeton MRI
experimental apparatus has been constructed to minimize the circulation by the use of independently controlled split endcaps \cite{kjg04,bsj06,jbsg06}. Nevertheless the jump of the rotation speed at the junction of the rings extends some distance into the bulk as a \emph{Stewartson layer} \cite{sk57,sk66}, however, the modification of the Stewartson layer by the axial magnetic field has to be studied. 

There has been research done on the MHD Ekman layers (or Ekman-Hartmann layers as they are sometimes called in the literatures) \citep{gb68,bl69,lb70,ld70,gp71,bc72,ga72}. However there has been little work aside from \citet{pv03} concerning the effect of finite differential rotation on Ekman layers, or on magnetized Stewartson layers in cylindrical geometry \cite{sk57,sk66} (\citet{hr94,hr96} did excellent work on magnetized Stewartson layers, though in spherical geometry). The latter issues remain poorly understood but play a big role in MRI experiments \cite{lgj07,sr07} and have potential importance in geophysics and fluid dynamics. \citet{hf04} discussed the purely hydrodynamic (unmagnetized) steady results with the assumption of infinitesimal differential rotation, or a very tiny Rossby number, while our paper discusses time-dependent solutions with finite differential rotation. \citet{sr07} presented results with finite differential rotation but without rings (similar to Sec. III of our paper), thus no Stewartson layer is present. That paper is a good contribution to Potsdam Rossendorf Magnetic Instability Experiment (PROMISE) and the discussion of Taylor-Dean flow is very insightful. This paper is one of the first to study magnetized Stewartson layers in cylindrical geometry. Understanding the role of the boundary layers, especially magnetized ones, is critical to the success of MRI experiments. 

It is known that
Ekman circulation is significantly modified when the
Elsasser number \cite{gp71} exceeds unity:
\begin{equation}
\label{elssasser}
\Lambda=B_{z0}^{2}/(8\pi\rho\eta\Omega)\gtrsim 1\; ,
\end{equation}
where $\Omega$ is the characteristic rotation frequency, which we take equal to $\Omega_2$. For gallium,
$\Lambda\approx2.5(B/{\rm Tesl a})^{2}(1000\;\unit{rpm}/\Omega)$.
Adopting the parameters used in the Princeton MRI experiment (Table.\ref{Ta_parameter}), $\Omega=\Omega_{2}=533\;\unit{rpm}$, $B=5000\;\unit{Gauss}$, immediately leads to $\Lambda\sim1.2$. In the PROMISE experiment \cite{sgg06}, $\Lambda\sim2.4$. Hence magnetic modifications to the Ekman layer should be significant in both experiments. 

\begin{figure}[!htp]
\begin{center}

  \scalebox{0.4}{\includegraphics{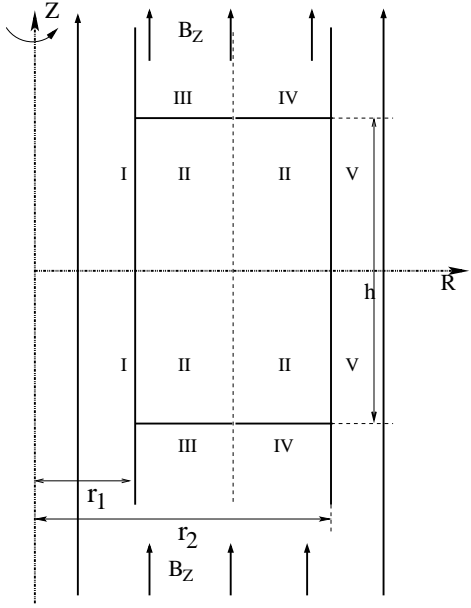}}
  \caption{\label{ekman_scheme}   Computational domain for studies of 
  magnetic Ekman layer. Region
  (I): Perfect conducting inner cylinder, angular velocity $\Omega_{1}$, infinitely long.
  (II): Liquid metal. (III): Perfectly insulating inner ring, $\Omega_{3}$, extending to infinity;  (IV): Perfectly insulating outer ring, $\Omega_{4}$, extending to infinity; (V): Perfectly conducting outer cylinder, $\Omega_{2}$, infinitely long. Thin dash line: the midplane. $B_{z}$ is the initial background vertical uniform magnetic field.}
\end{center}
\end{figure}

Here we report nonlinear simulations with the \emph{ZEUS-2D} code \cite{sn921,sn922}, which is a time-explicit, compressible, astrophysical ideal magnetohydrodynamics (MHD) two-dimensional code, to which we have added viscosity and resistivity (with subcycling to reduce
the cost of the induction equation) for axisymmetric flows in cylindrical coordinates $(r,\varphi,z)$ \cite{lgj06}. The simulation domain mimics the Princeton MRI experiment (see Table~\ref{Ta_parameter} and Fig.~\ref{ekman_scheme}) except where stated explicitly. 
The code adopts the magnetic boundary conditions introduced in Sec.~II.D of \citet{lghj06} (not the commonly used vertically pseudo-vacuum boundary conditions). All real flows are actually compressible; in
an ideal gas of fixed total volume, density changes generally scale
$\sim M^2$ when Mach number $M=V_{\rm flow}/V_{\rm sound}< 1$.
Incompressibility is an idealization in the limit $M\to0$. An isothermal equation of state has been used with a sound speed chosen so that the maximum of $M\le 1/4$. The techniques used here have been benchmarked analytically and compared with other codes in \citet{lgj06,lghj06} and verified experimentally in \citet{bsj06,lgj07}. Note that in the simulations the magnetic diffusivity $\eta$ is fixed to the experimental value $\eta\sim2,000\;\unit{cm^{2}\;s^{-1}}$ (Table.\ref{Ta_parameter}), however the kinematic viscosity is varied for the purpose of extrapolation.  Also as demonstrated by \citet{gj02}, the viscosity of liquid metals is so small as to be almost irrelevant to MRI, at least in the linear regime.

\begin{figure}[!htp]
\begin{center}

  \scalebox{0.4}{\includegraphics{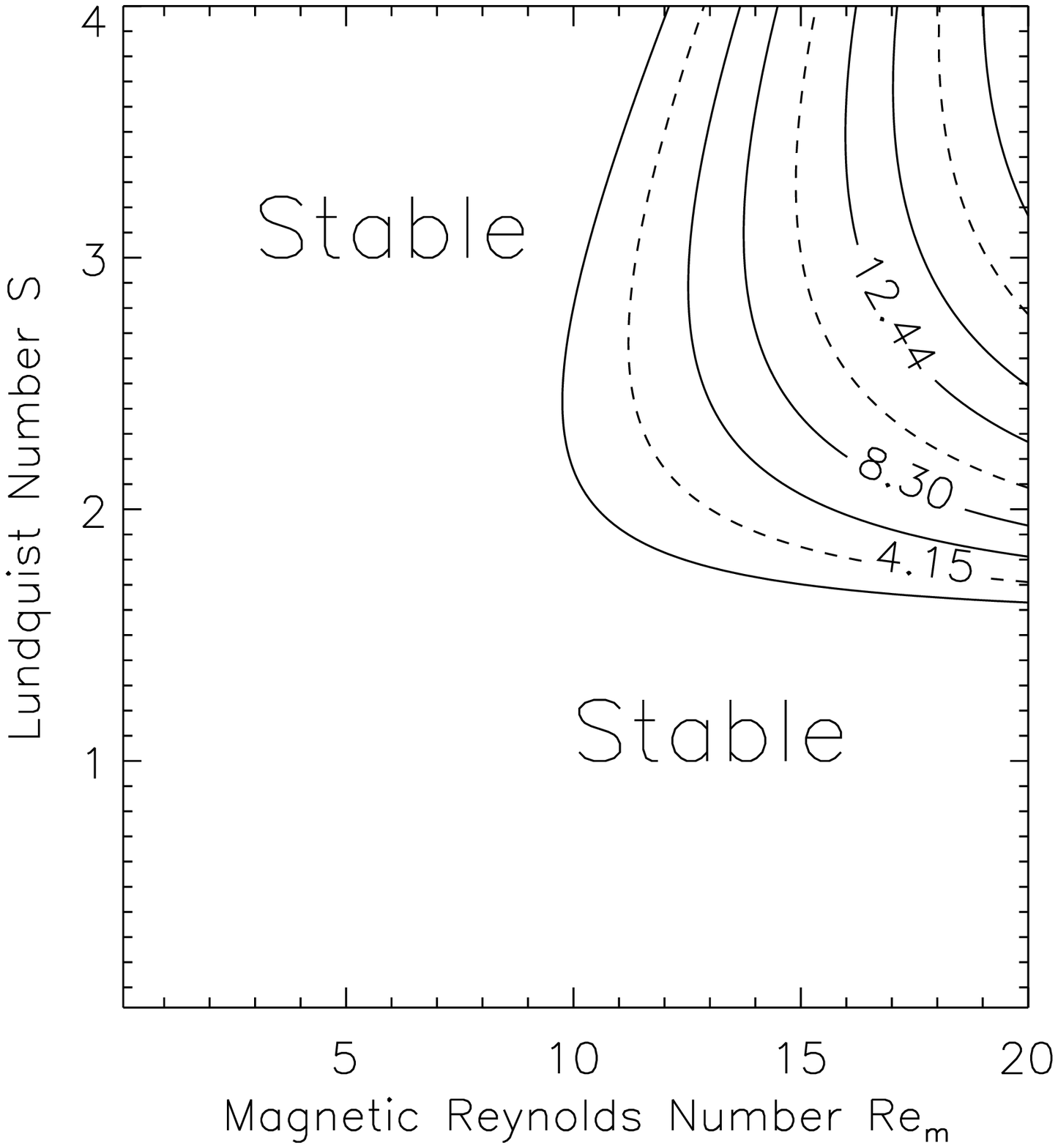}}
  \caption{\label{marginal}  MRI linear growth rates ($\gamma\;\unit{s^{-1}}$) as a function of magnetic Reynolds number $\Rm$ and Lundquist number $S$ with dimensions as in Table~\ref{Ta_parameter}: $r_1=7.1\;\unit{cm}$, $r_2=20.3\;\unit{cm}$ and $h=27.9\;\unit{cm}$ and $\mu=\Omega_2/\Omega_1=0.13325$. The system is MRI stable if $\Rm\lesssim10$ regardless of $S$.  }
\end{center}
\end{figure}

In this present work, since we concentrate on magnetic Ekman and Stewartson layers, the rotation speed profile is chosen so that the system is MRI stable. We have found the MRI linear growth rates ($\gamma\;\unit{s^{-1}}$) as a function of magnetic Reynolds number $\Rm$ and Lundquist number $S$ (Fig.~\ref{marginal}) from  a WKB analysis \cite{jgk01} with the same dimensions as in Table~\ref{Ta_parameter} and $\mu=\Omega_2/\Omega_1=0.13325$, which shows that for $\Rm\lesssim10$ the system should be MRI stable. All simulations presented in this paper are in this regime.

This paper is organized as follows. 
We start in Sec.~\ref{infinitesimal} by reviewing a one-dimensional approximation to magnetized Ekman circulation above an infinite, uniformly rotating boundary.  Two-dimensional effects are introduced in Sec.~\ref{sec:finite}, but still with rigid endcaps.  In Sec.~\ref{split}, we divide each endcap into two independently rotating rings as in the Princeton MRI experiment, and study the dependence of the resulting Stewartson layer on the Reynolds and Elsasser numbers.  Implications for the Princeton MRI experiment are discussed in Sec.~\ref{implications}.

\begin{table}
 \begin{center}
 \begin{tabular}{| c | c |} \hline
 \multicolumn{2}{| c |}{ Dimensions} \\ \hline
 $r_{1}=7.1\;\unit{cm}$ & $r_{2}=20.3\;\unit{cm}$ \\ \hline
  \multicolumn{2}{| c |} {$h=27.9\;\unit{cm}$}  \\ \hline
 \multicolumn{2}{| c |}{Material Property} \\ \hline 
 $\rho\approx6.0 \;\unit{g cm^{-3}}$ & $\eta\approx2.0\times10^{3}\;\unit{cm^{2} s^{-1}}$ \\ \hline
 \multicolumn{2}{| c |}{ Full Speed Run} \\ \hline
 $\Omega_{1}/2\pi=4000\;\unit{rpm}$ & $\Omega_{2}/2\pi=533\;\unit{rpm}$ \\ \hline
 $\Omega_{3}/2\pi=1820\;\unit{rpm}$ & $\Omega_{4}/2\pi=650\;\unit{rpm}$ \\ \hline
  \multicolumn{2}{| c |}{ Rotation Profile used in Sec.~\ref{sec:finite}} \\ \hline
 $\Omega_{1}/2\pi=500\;\unit{rpm}$ & $\Omega_{2}/2\pi=66.625\;\unit{rpm}$ \\ \hline
 $\Omega_{3}/2\pi=66.625\;\unit{rpm}$ & $\Omega_{4}/2\pi=66.625\;\unit{rpm}$ \\ \hline 
   \multicolumn{2}{| c |}{ Rotation Profile used in Sec.~\ref{split}} \\ \hline
 $\Omega_{1}/2\pi=500\;\unit{rpm}$ & $\Omega_{2}/2\pi=66.625\;\unit{rpm}$ \\ \hline
 $\Omega_{3}/2\pi=227.5\;\unit{rpm}$ & $\Omega_{4}/2\pi=81.25\;\unit{rpm}$ \\ \hline 
 \end{tabular}
  \caption{Parameters used in the simulations}\label{Ta_parameter}
  \end{center}
 \end{table}

\section{Standard Magnetic Ekman Layer with near-uniform rotation}\label{infinitesimal}

We begin with a problem considered by \citet{gb68}.
The problem treated consists of an incompressible, viscous and resistive fluid above an infinite, flat and insulating boundary that rotates at angular velocity $\boldsymbol{\Omega}=\Omega\boldsymbol{e}_{z}$. Far from the boundary, the fluid rotates uniformly at $\Omega^{'}=\Omega(1+\epsilon)$. A uniform magnetic field aligned with the rotation axis is imposed. In the analysis of \citet{gb68}, an expansion in powers of $\epsilon$, together with von K\'arm\'an similarity \cite{vkt30,gs37}, leads to a solution that is exact to first order in $\epsilon$. In the limit that $\epsilon\ll 1$, increasing $\Lambda$ results in a continuous transition between pure Ekman flow and a rotating analog of Hartmann flow. 

Here we sketch a modified steady state WKB analysis rather than an expansion in the von K\'arm\'an similarity variables used by \citet{gb68}. With the $t$ and $r$ dependence factored out, the
linearized equations of motion reduce to inhomogeneous ordinary
differential equations with coefficients independent of $z$.
Elementary homogeneous solutions of these equations exist with exponential dependence on
$z$; however, since there is an insulating endplate at $z=0$, the wavenumber $k_{n}$ may be complex, and
the final solution can be a linear combination of the elementary modes for different $k_n$ and one particular solution that matches the flow far from the boundary.
The vertical magnetic boundary conditions require the fields to match
onto a vacuum solution at the end plate.

We seek a mode of the form
\begin{equation}
\label{eqn16}
 [v_{r},v_{\varphi},B_{r},B_{\varphi}]^{T}=
~[0,V_{\infty},0,0]^{T}+\sum\limits_{n=1}^8 C_n
[v_{r,n},v_{\varphi,n},B_{r,n},B_{\varphi,n}]^{T}
\exp(ik_{n}z).
\end{equation}
The first column vector on the right-hand side is the particular solution, which satisfies the boundary conditions at $z=\infty$ but not at $z=0$, where $V_{\infty}$ is the velocity far away from the end plate. Each term in the sum above is the elementary solution corresponding
to a particular wavenumber $k_{n}$, with $(v_{r,n},\dots,B_{\varphi,n})^{T}$ a 4-component
column vector; these elementary solutions are superposed with constant weights
$\{C_{n}\}$, which must be chosen to satisfy the boundary condition. The $8$ values of the wavenumber $\{k_{n}\}$ are the roots of the steady-state dispersion relation:

\begin{equation}
k^{4}[(\eta\nu)^{2}k^{4}+2\eta\nu V_{A}^{2}k^{2}+(V_{A}^{4}+4\Omega^{2}\eta^{2})]=0\, ,
\end{equation} 
which follows from the linearized, homogeneous and axisymmetric Navier-Stokes and induction equations. 
Only the four nonzero roots of this equation are of interest since they determine the boundary-layer thickness. The eigenmodes corresponding to $k=0$ would modify the interior flow ($z\to\infty$). 
The four nonzero roots are
\begin{equation}
\label{eqn17}
k^{2}=\frac{V_{A}^{2}}{\eta\nu}\pm\frac{2\Omega i}{\nu}\, .
\end{equation}
The two ``acceptable" nonzero roots of Eq.~\ref{eqn17}, satisfying the boundary conditions, 
are $k_{\pm}=-(k_{R}\pm ik_{I})$, where $k_{R}=\delta^{-1}$ as given by Eqn~\ref{eqn18}, so that
\begin{eqnarray}
\label{gb_vx}
v_{r} & = & -V_{\infty}e^{-k_{R}z}\sin k_{I}z\, , \\
\label{gb_vy}
v_{\varphi} & = & V_{\infty}(1-e^{-k_{R}z}\cos k_{I}z)\, , 
\end{eqnarray}
where $k_{I}$ is related to $k_{R}$ by $k_{I}/k_{R}=\sqrt{1+\Lambda^{2}}-\Lambda$.
Thus
\begin{equation}
\label{eqn18}
\delta=\delta_{E}\frac{1}{\sqrt{\sqrt{\Lambda^{2}+1}+\Lambda}}\approx \delta_{E}\times
\begin{cases}
1-\Lambda/2 &\text{if $\Lambda \ll 1$}\, ; \\
1/\sqrt{2\Lambda} &\text{ if $\Lambda \gg 1$}\, .
\end{cases}
\end{equation}
Here $\delta_{E}=\sqrt{\nu/\Omega}$ is the purely hydrodynamical Ekman-layer thickness with near-uniform rotation. Notably, the Elsasser number $\Lambda$ (Eq.~\ref{elssasser}) has nothing to do with $\nu$. Hence even if the boundary layer were turbulent, with an effective turbulent viscosity $\nu_{T}$ and thickness increased by $O[(\nu_{T}/\nu)^{1/2}]$, the magnetic field would be at least as consequential as in the laminar case. This assumes that turbulent magnetic diffusivity is negligible, as one might expect since the laminar value is large enough.
One expects that $\Lambda\gg1$ probably results in a more stable layer and pushes the onset of turbulence to larger Reynolds numbers. 
In the limit $\Lambda\to\infty$, the thickness $\delta\to\sqrt{\nu\eta}/V_{A}$: this is the Hartmann-layer thickness, which does not depend upon $\Omega$.

\begin{figure}[!htp]
\begin{center}

\scalebox{0.4}{\includegraphics{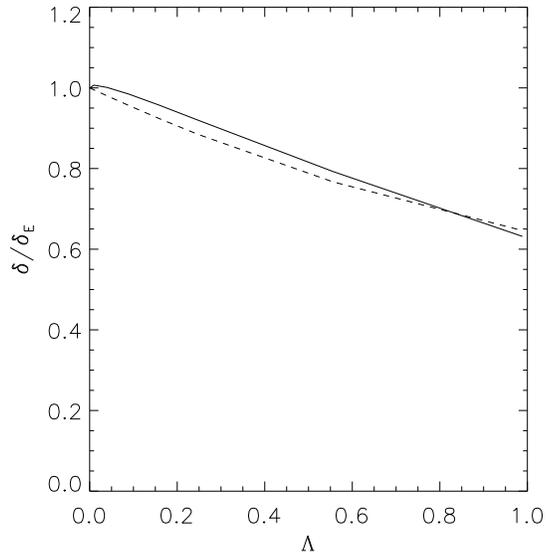}}
\caption{\label{standard_ekman} The thickness of the Ekman layer $\delta$ versus Elsasser Number $\Lambda$ for $Re=1600$, $\Rm=5$. $\Omega_{1}/2\pi=1000\;\unit{rpm}$, $\Omega_{2}/2\pi=1000\;\unit{rpm}$, $\Omega_{3}/2\pi=1010\;\unit{rpm}$, $\Omega_{4}/2\pi=1010\;\unit{rpm}$. $r_{1}=15\;\unit{cm}$, $r_{2}=35\;\unit{cm}$ and $h=20\;\unit{cm}$. The data is measured at $r=(r_{1}+r_{2})/2=20\;\unit{cm}$. The dashed line is the theoretical result. The solid line is the one obtained from modified ZEUS-2D simulations. We have chosen larger $r_1$ and $r_2$ than the ones of the Princeton MRI experiment to minimize the curvilinear effects and larger $h\gg10\delta_E$ to minimize the influence of the top endcap. }
\end{center}
\end{figure}

The above theoretical results have been used to benchmark our code (Fig.~\ref{standard_ekman}). The thickness of the Ekman layer $\delta$ is the reciprocal of $k_R$, which is deduced by fitting the simulated data at $r_d=(r_1+r_2)/2$ using Eq.~\ref{gb_vx}.
The results agree well with the theoretical prediction (Eq.~\ref{eqn18}).

\section{Magnetic Ekman Layer with differential rotation: (I) endplates corotating with outer cylinder}\label{sec:finite}

In contrast with  the idealized case in Sec.~\ref{infinitesimal}, most Taylor-Couette experiments have a two-dimensional circulation driven by differences in the rotation of the inner and outer cylinders and the endcaps.  In this section, we take the end plates to corotate with the outer cylinder, \emph{i.e.}, $\Omega_3=\Omega_4=\Omega_2$.

The Reynolds number based on  the Ekman layer thickness is \cite{pv03}:
\begin{equation}
Re_{\delta}\approx\frac{r_{2}\Omega_{2}\delta}{\nu}\approx Re^{1/2}\sim3\times10^{3}\, ,
\end{equation}
for full-speed runs of the Princeton MRI experiment (Table~\ref{Ta_parameter}). The Ekman layer with uniform rotation as in Sec.~\ref{infinitesimal} has two known instabilities, viscous and inflection point instabilities, both of which are axisymmetric. The viscous instability owes its existence to the perturbed Coriolis force \cite{ld66} while the inflection point instability is of the inviscid type. From Fig.~3 of \citet{gp71}, the critical Reynolds number of the viscous and inflection point instabilities associated with this  Ekman-Hartmann layer is in the range: $100\lesssim Re_{\delta c} \lesssim1000$ for $\Lambda\approx1$, at least for cases of near-uniform rotation as in Sec.~\ref{infinitesimal}. Thus the boundary layer is turbulent for the full-speed runs of the Princeton MRI experiment. However in the simulations, the Reynolds number in the bulk is taken to be $6400$, thus $Re_{\delta}=80$, so that the boundary layer is laminar. Our discussion below is grounded on the equations of laminar flows. 
The magnetic Reynolds number in the boundary layer based on the thickness of the Ekman layer is defined as \cite{pv03}:
\begin{equation}
Rm_{\delta}=\frac{\delta U_{0}}{\eta}\approx\frac{\Rm}{\sqrt{Re}}\, ,
\end{equation} 
where $U_{0}$ is a characteristic speed. 
For $Re=6400$ and $\Rm=2.5$ as in the simulations, $Rm_{\delta}\approx3.125\times10^{-2}$. Because $Rm_{\delta}\ll1$ and $|\omega=(\Omega-\Omega_{2})/\Omega_{2}|\ll1$, 
the quantity $1+(1/2)r_*d\omega/dr_{*}=1/(2\Omega r)d(r^{2}\Omega)/dr$, which is the ratio of vorticity to rotation ($r_*$ is the radius normalized by $r_2$), is $\approx a/\Omega_2>0$.
The solution decays with oscillation as $z\to\infty$, as for an unmagnetized Ekman layer.
The modified Ekman layer thickness $\delta$ is given by \cite{pv03}:
\begin{equation}
\label{pv_thickness}
\delta=\delta_{E}(\alpha_{1}^{'})^{-1}=\delta_{E}\frac{1}{\sqrt{\sqrt{\Lambda^{2}+1+\frac{1}{2}r_{*}\frac{d \omega}{dr_{*}}}+\Lambda}}\, .
\end{equation}
Eq.~\ref{pv_thickness} reproduces Eq.~\ref{eqn18} if there is no differential rotation ($d\omega/dr_{*}\to0$). Therefore \emph{the strong magnetic field causes the Ekman layer to become thinner even with differential rotation.} 
It is worth emphasizing that the above derivation is based on a first order expansion in $\omega\ll 1$. 

Our simulations with the parameters of Table~\ref{Ta_parameter} approach the regime n of the above linear theory except that: (1) the radial boundary condition is conducting rather than insulating (a magnetic Ekman layer with fully insulating boundaries on all sides is the next step for this problem and will be included in a forthcoming paper); (2) the flow profile far away from the end plate differs from the ideal Couette profile, though not greatly; (3) $|(\Omega-\Omega_{2})/\Omega_{2}|\ll1$ is not satisfied except near the outer cylinder.

The simulations are analyzed at $r_{d}=(r_{1}+r_{2})/2$ to minimize two-dimensional effects due to the radial boundary. Since $|(\Omega(r_d)-\Omega_{2})/\Omega_{2}|=1.08$ is not small, some nonlinear effects neglected in the above linear analysis could be important.

\begin{figure}
\begin{center}

\scalebox{0.4}{\includegraphics{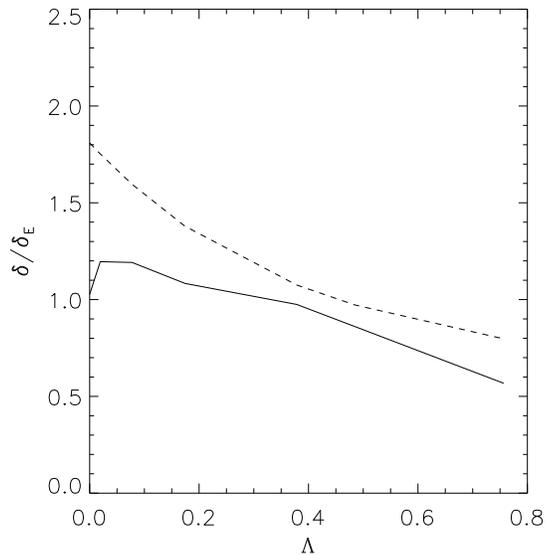}}
\caption{\label{finite_ekman} The thickness of the Ekman layer $\delta$ versus Elssaser Number $\Lambda$ for $Re=6400$, $\Rm=2.5$. Parameters as in Table~\ref{Ta_parameter}. The data are measured at $r=(r_{1}+r_{2})/2=13.7\;\unit{cm}$. The dashed line is from the linear analysis (Eq.~\ref{pv_thickness}). The solid line is obtained from modified ZEUS-2D simulations. }
\end{center}
\end{figure}

From Fig.~\ref{finite_ekman}, we confirm that the axial magnetic field does reduce the Ekman layer thickness. The finite differential rotation cannot be neglected and modifies the linear Ekman layer, which is seen from the unmagnetized case, \emph{i.e.} $\Lambda=0$: The theoretical result predicts that the thickness of the Ekman layer with finite differential rotation is larger than the thickness of the Ekman layer with uniform rotation; in the numerical results, this is less clear.  Though the simulated curve does not match the theoretical result very well, the agreement is as good as might be expected when a theory based on $\omega\ll1$ is applied to simulations at $\omega\sim1$.

\begin{figure}

\scalebox{0.4}{\includegraphics{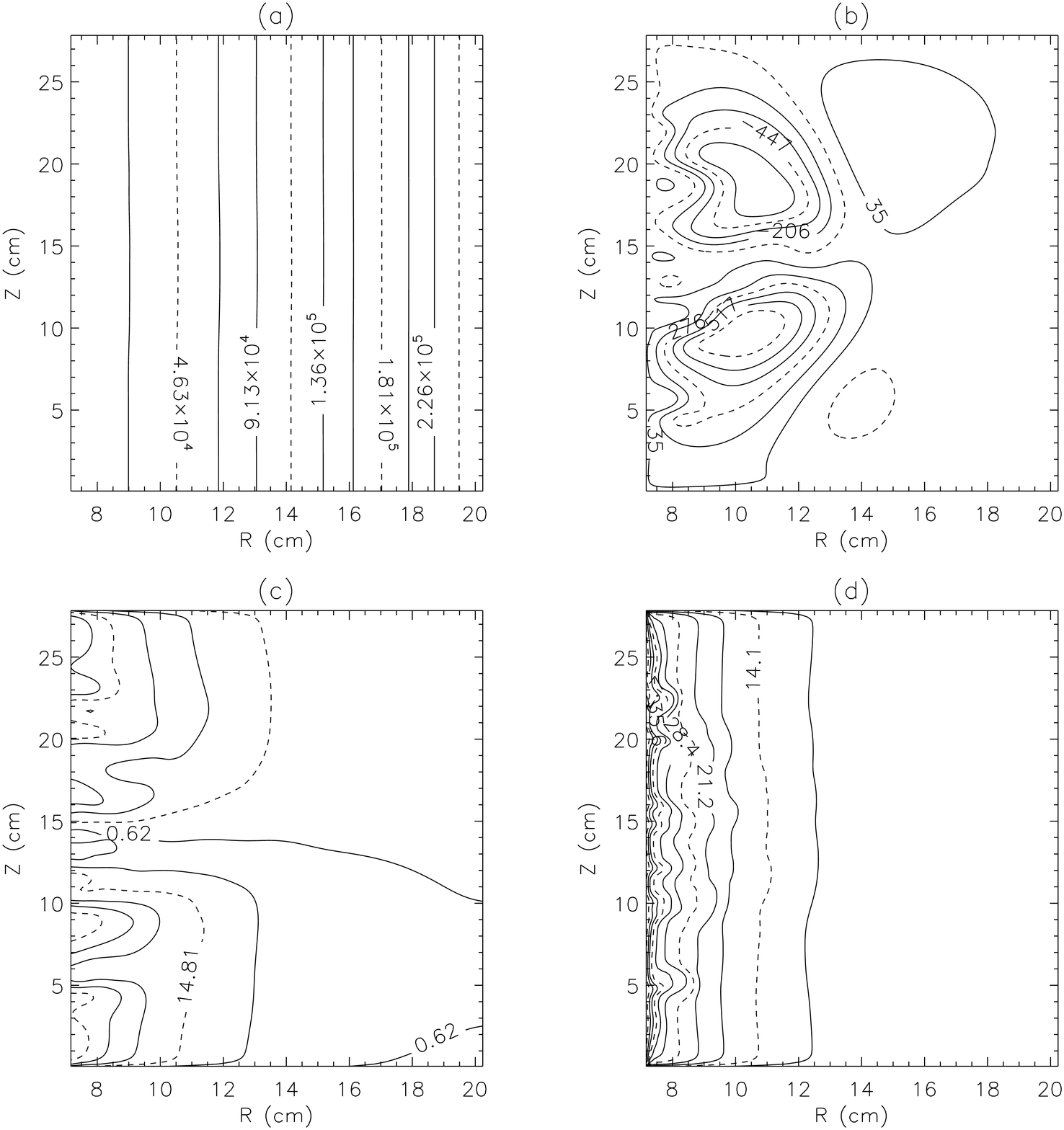}}
\caption{\label{finite_ekman_final_pattern} Contour plots of final-state
velocities and fields with uniformly rotating endcaps.  $Re=6400$, $\Rm=2.5$ with $B_{z0}=1500\;\unit{Gauss}$ ($\Lambda=1.09$). Parameters as in Table~\ref{Ta_parameter}.  (a) Poloidal flux
function $\Psi \unit{(Gauss\,cm^{2})}$ (b) Poloidal stream function
$\Phi \unit{(cm^{2}s^{-1})}$ (c) toroidal field
$B_{\varphi} \unit{(Gauss)}$ (d) angular velocity
$\Omega\equiv r^{-1}V_{\varphi}\unit{(rad\,s^{-1})}$.}
\end{figure}

For $Re=6400$, 
the final state is not steady even after at least five Ekman times $\tau_E=h/(\nu\bar{\kappa}/2)^{1/2}$. Given a finite differential rotation, it is more appropriate to estimate Ekman time $\tau_E$ via the epicyclic frequency,  
\[
\bar{\kappa}=2\left((r_{2}^{4}\Omega_{2}^{2}-r_{1}^{4}\Omega_{1}^{2})/(r_{2}^{4}-r_{1}^{4})\right)^{1/2}\; ,
\] 
which is the maximum frequency of small axisymmetric inertial oscillations inside the inviscid fluid. Typical (instantaneous) flow and
field patterns are shown in
Fig.~\ref{finite_ekman_final_pattern}.  The poloidal flux
and stream functions are defined so that
\begin{equation}\label{pfuncs}
\boldsymbol{V}_P\equiv
V_r\boldsymbol{e}_r+V_z\boldsymbol{e}_z=r^{-1}\boldsymbol{e}_\varphi
\boldsymbol{\times\nabla}\Phi,\qquad \boldsymbol{B}_P\equiv
B_r\boldsymbol{e}_r+B_z\boldsymbol{e}_z=r^{-1}\boldsymbol{e}_\varphi
\boldsymbol{\times\nabla}\Psi,
\end{equation}
which imply $\boldsymbol{\nabla\cdot V}_P=0$
and $\boldsymbol{\nabla\cdot B}_P=0$. Two Ekman cells are clearly visible. The flapping ``jet" at the midplane due to the Ekman circulation breaks the vertical refelction symmetry of the system, resulting in a chaotic region around the midplane \cite{mbjl07}. The poloidal flow circulation and toroidal field are small compared to the background toroidal flow and initial axial field respectively,
\[
{\rm max}\frac{v_{r}}{r_1\Omega_1}\lesssim13\%\;,\quad{\rm max}\frac{B_{\varphi}}{B_{z0}}\lesssim3\%\,.
\]

\begin{figure}[!htp]
\begin{center}

\scalebox{0.4}{\includegraphics{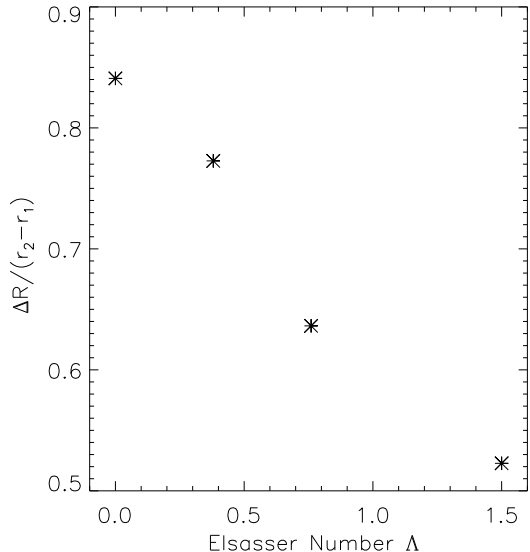}}
\caption{\label{solid} $\Delta R/(r_2-r_1)$ \emph{v.s.} $\Lambda$. $\Delta R$ is the radial width of the poloidal circulating region. $\Rm=2.5$, $Re=6400$ with endplates corotating with outer cylinder. Parameters as in Table~\ref{Ta_parameter}. From the radial profile of the azimuthal velocity $v_{\varphi}$, $\Delta R$ is the  radial gap between the inner cylinder and the place where the solid body rotation starts. Note that in the simulations the magnetic diffusivity $\eta$ is fixed to $\eta\sim2,000\;\unit{cm^{2}\;s^{-1}}$, however the imposed axial magnetic field $B_{z0}$ is varied.}
\end{center}
\end{figure}

The most noticeable feature of the final state of the magnetic Ekman circulation is the presence of an area of solid body rotation near the outer cylinder (Fig.~\ref{finite_ekman_final_pattern} (d)) as in \citet{hr94}. This area increases with the Elsasser number: the strong axial magnetic field squeezes the dynamically active area (Ekman cells) toward the inner cylinder (Fig.~\ref{solid}). When $\Lambda=1.5$, almost half of the liquid metal is rotating with the outer cylinder. This is due at least in part to the following two effects: (1) Larger axial magnetic fields suppress the Ekman circulation more thoroughly; (2) The axial Hartmann current turns towards the radial direction near the midplane, and couples with the axial magnetic field to produce an azimuthal Lorentz force, which tends to reduce the azimuthal velocity shear $\partial \Omega/\partial z$. Both effects reinforce Taylor-Proudman theorem near the inner cylinder, where there is a large velocity shear between the bulk flow and the end plate. 

\section{Magnetic Ekman layer with differential rotation: (II) end plates split into two rings}\label{split}

\begin{figure}[!htp]

\scalebox{0.4}{\includegraphics{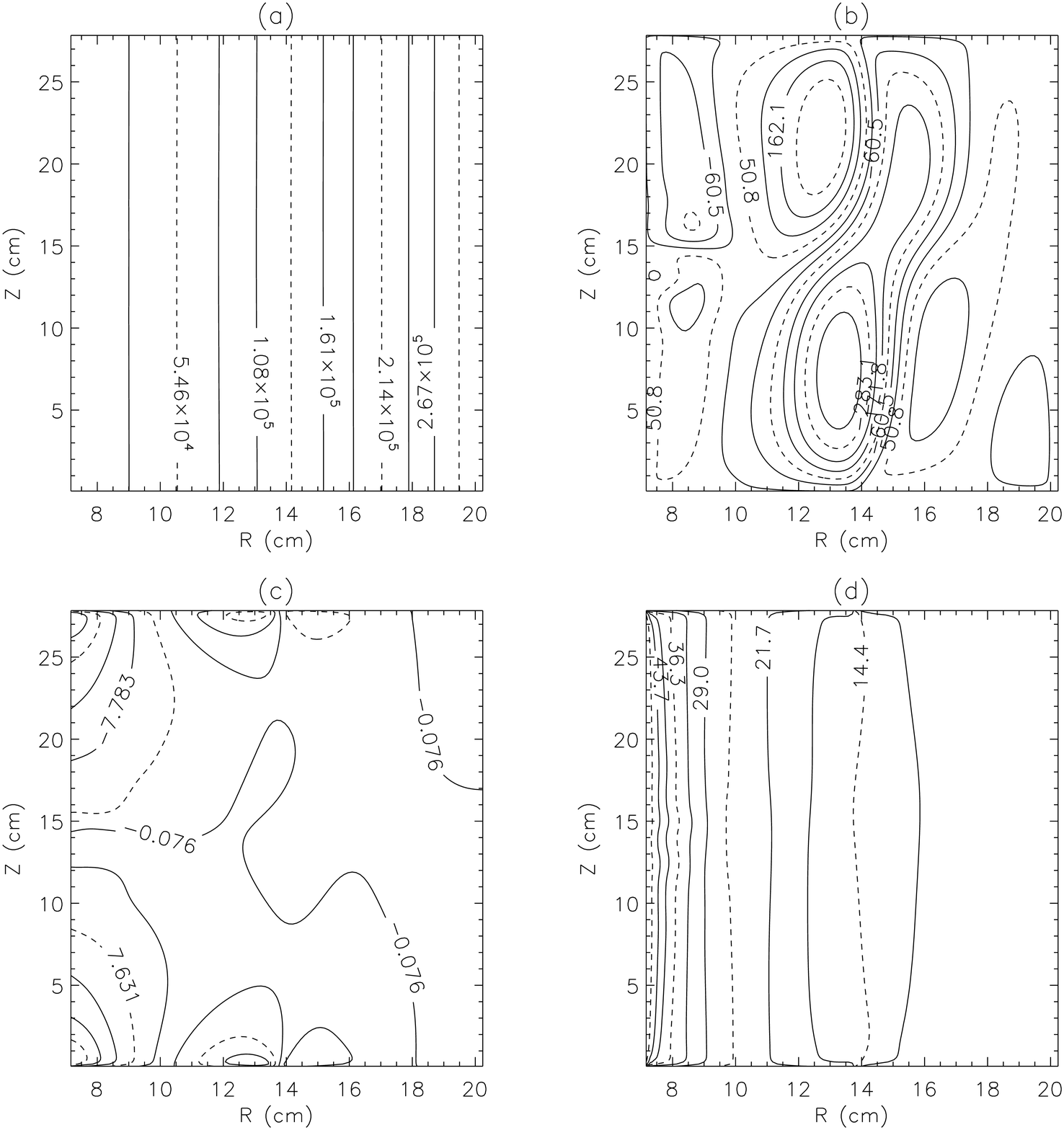}}
\caption{\label{ring2_ekman_final_pattern} Like Fig.~\ref{finite_ekman_final_pattern},
but with two differential rotating rings and $\Lambda=1.5$. Parameters as in Table~\ref{Ta_parameter}. The Stewartson layer is located between
the rings at $r_d=(r_1+r_2)/2=13.7\;\unit{cm}$ and breaks the two big Ekman cells into eight cells.}
\end{figure}

We have brought the computation closer to the experimental conditions by making the endcaps consist of two independently rotating rings as in Fig.~\ref{ekman_scheme} and Table~\ref{Ta_parameter}. The junction between these two rings lies at $r_d=(r_1+r_2)/2=13.7\;\unit{cm}$.
For $Re=6400$, the final state is not steady. Typical (instantaneous) flow and
field patterns are shown in
Fig.~\ref{ring2_ekman_final_pattern}. Two flapping ``jets" due to the unsteady Stewartson layer, emanating from the junction of the rings at both endcaps, leads to a chaotic region localized  there (Fig.~\ref{ring2_ekman_final_pattern} (b)), which is different from the case in Sec.~\ref{sec:finite}, in which the unsteady region is mainly near the midplane. The poloidal flow circulation and toroidal field are also small compared to the background toroidal flow and initial axial field respectively,
\[
{\rm max}\frac{v_{r}}{r_1\Omega_1}\lesssim4.3\%\;,\quad{\rm max}\frac{B_{\varphi}}{B_{z0}}\lesssim1.3\%.
\]
These ratios are smaller than the ones discussed in Sec.~\ref{sec:finite}, which implies that the Ekman suction is reduced by splitting the endcaps into two differentially rotating rings.

The following observations can be made from Fig.~\ref{ring2_ekman_final_pattern}. With rings the Stewartson layer is more apparent than Ekman circulation. The split endcaps break the two big Ekman cells found in Sec.~\ref{sec:finite} (Fig.~\ref{ring2_ekman_final_pattern} (b)). The four cells at intermediate radii are straightforward consequences of the Stewartson layer as discussed below. The direction of the circulation of the bottom four cells is opposite to the direction of the circulation of the corresponding upper cells. Hereafter we focus only on the upper half of the flow. 

\begin{figure}[!htp]
\begin{center}

\scalebox{0.4}{\includegraphics{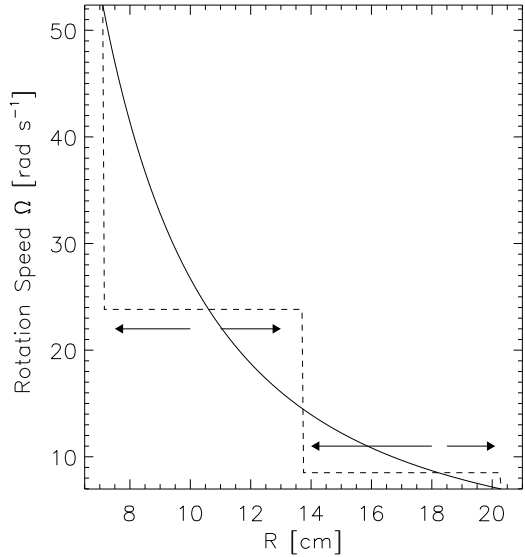}}
\caption{\label{break} Two independently rotating rings generate $8$ cells. solid line, ideal Couette state; dashed line, rotation profile at the endcaps. Arrows indicate the radial flow directions near the endcaps.}
\end{center}
\end{figure}

The increase of the number of Ekman cells can be understood from Fig.~\ref{break}. The direction of the residual Ekman flow depends upon the angular velocity of the boundary relative to the interior, thus resulting in anti-clock-wise normal Ekman cells at $r\lesssim10.6\;\unit{cm}$ and $13.7\;\unit{cm}\lesssim r \lesssim18.2\;\unit{cm}$ and clock-wise abnormal Ekman cells elsewhere.

\begin{figure}[!htp]
\subfigure{\scalebox{0.4}{\includegraphics{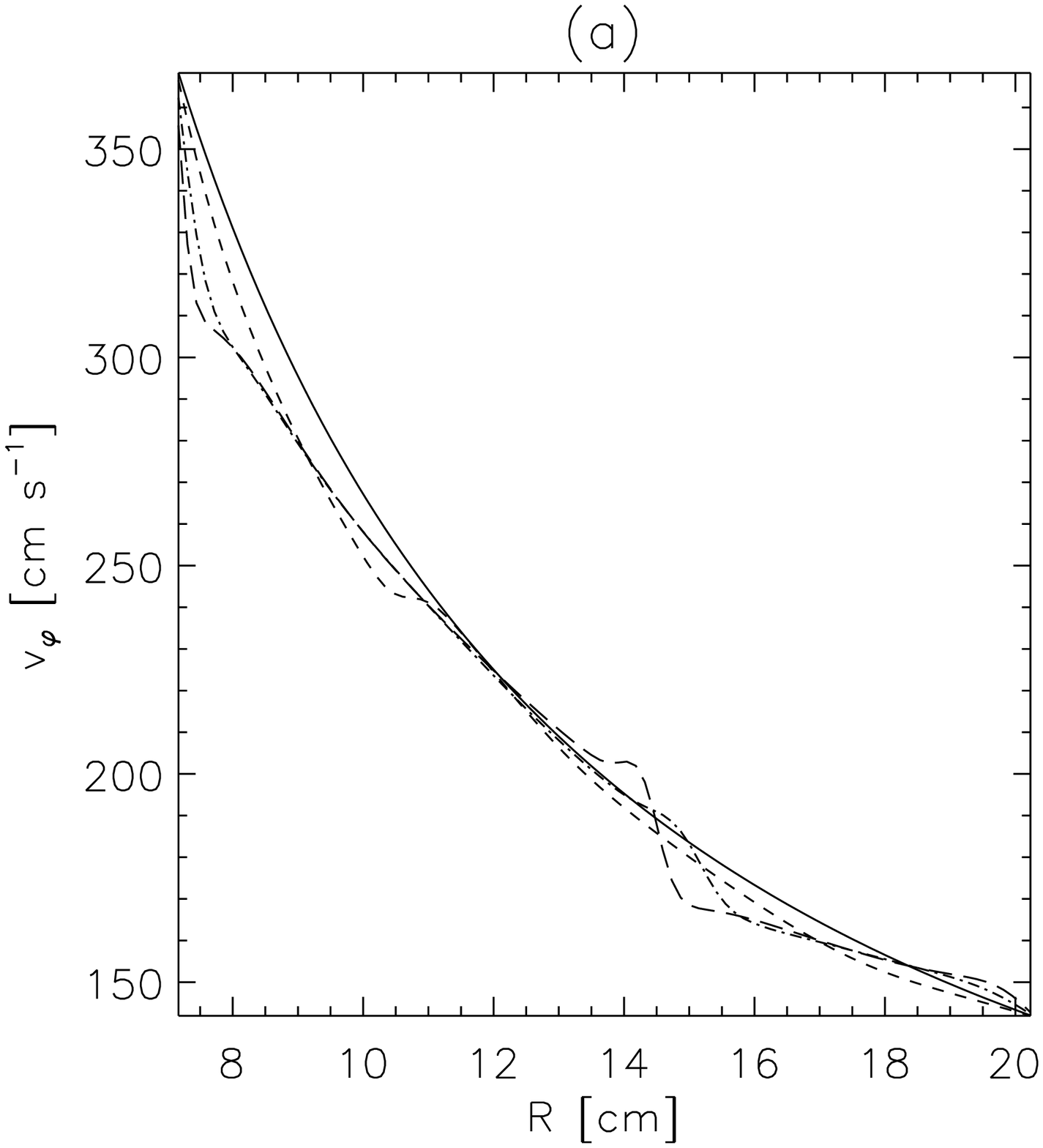}}}$\,$
\subfigure{\scalebox{0.4}{\includegraphics{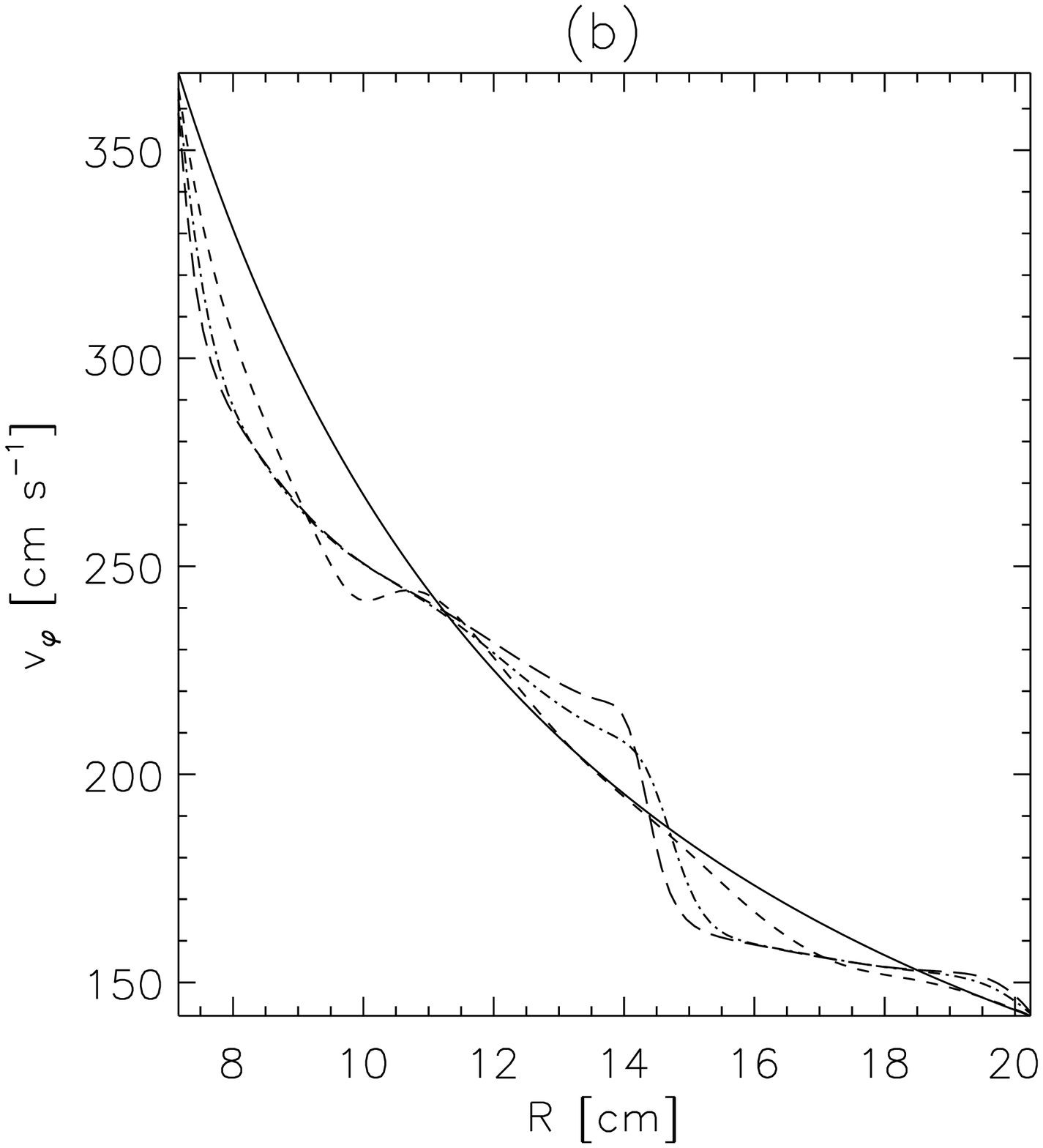}}}
\caption{\label{ring2_ekman_vt} Azimuthal velocity $v_{\varphi}\;\unit{cm\;\s^{-1}}$ versus radius $r$ at different heights with $\Rm=2.5$, $Re=6400$, and the endcaps divided into two rings at $r_d=(r_1+r_2)/2=13.7\;\unit{cm}$. Parameters as in Table~\ref{Ta_parameter}. Solid line, ideal Couette state; long dashes, $z=1.33\;\unit{cm}$ (relative to the bottom endcap); dash dot, $z=2.79\;\unit{cm}$; short dashes, $z=13.95\;\unit{cm}$. (a): $\Lambda=0.38$; (b): $\Lambda=1.5$. }
\end{figure}

The magnetic field tends to reduce fluctuations in the final state at high Reynolds number.
The Stewartson layer becomes more prominent with increasing $\Lambda$ at fixed $Re$ (Fig.~\ref{ring2_ekman_vt}). This can be understood by considering the influence of the magnetic field on the stability of the Stewartson layer. The addition of an axial magnetic field (in the MRI stable regime) resists shear along the magnetic field lines and elongates the cells vertically so that they penetrate deeply into the fluid. More details are given in Sec.~\ref{implications}.

\begin{figure}[!htp]
\begin{center}

\scalebox{0.4}{\includegraphics{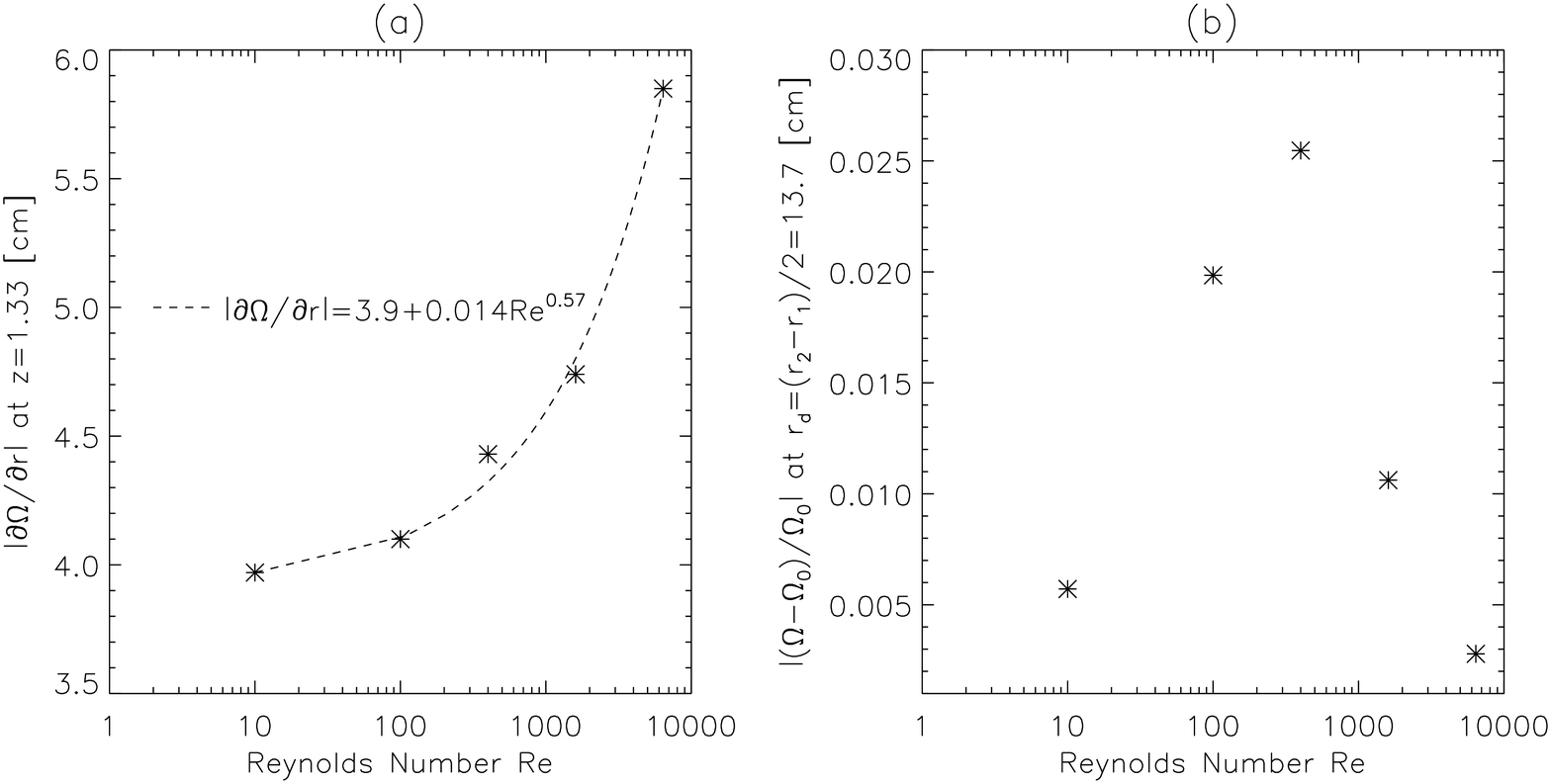}}
\caption{\label{zeta} (a) Time-averaged $|\partial\Omega/\partial r|$ of the Stewartson layer at $z=1.33\;\unit{cm}$ \emph{v.s.} Reynolds number $Re$. (b) Time-averaged $|[\Omega-\Omega_0(r_d)]/\Omega_0(r_d)|$ with $r_d=(r_1+r_2)/2=13.7\;\unit{cm}$ at the midplane. \emph{v.s.} $Re$, where $\Omega_0$ is the ideal Couette state defined in Eq.~\ref{couette}.  $\Rm=2.5$, $\Lambda=1.5$, and the endcaps divided into two rings at $r_d=(r_1+r_2)/2=13.7\;\unit{cm}$. Parameters as in Table~\ref{Ta_parameter}. Note that in the simulations the rotation speed profile is fixed to $\Omega_{1}/2\pi=500\;\unit{rpm}$, $\Omega_{2}/2\pi=66.625\;\unit{rpm}$, $\Omega_{3}/2\pi=227.5\;\unit{rpm}$, $\Omega_{4}/2\pi=81.25\;\unit{rpm}$, however the kinematic viscosity $\nu$ is varied.}
\end{center}
\end{figure}

On the other hand, the Stewartson layer also becomes sharper as $Re$ increases at fixed $\Lambda$ 
From Fig.~\ref{zeta} (a), we infer the following scaling law:
\[
|\frac{\partial\Omega}{\partial r}|=3.9+0.014Re^{0.57}\,.
\]
This is somewhat consistent with the one-dimensional analyses of a purely hydrodynamic Stewartson layer, which show that a Stewartson layer consists of nested layers of outer thickness $E^{1/4}$ and inner thickness $E^{1/3}$ \cite{sk66}, where $E=1/Re$ is the Ekman number. Considering the idealizations used in the analyses \cite{sk66} and the complications in our two-dimensional simulations, the agreement is as good as might be expected.

We also observe that the Stewartson layer penetrates more deeply into the bulk flow with larger $Re$ at fixed $\Lambda$, at least for $Re\lesssim400$ and $\Lambda=1.5$ in axisymmetry. This can be seen 
from Fig.~\ref{zeta} (b).
The profiles deviate from the ideal Couette state more with larger $Re$ ($Re\lesssim400$). However at even higher Reynolds number (\emph{i.e.}, $Re>400$), the Stewartson layer develops axisymmetric MRI/centrifugal instabilities at large axial wave numbers, which upon saturating result in a less prominent Stewartson layer, as Fig.~\ref{zeta} (b) ($Re\gtrsim400$).
The layer is more localized near the endcaps in these cases. More details are given in Sec.~\ref{implications}.

\section{Discussion}\label{implications}

Purely hydrodynamical (\emph{i.e.}, $\Lambda=0$) experimental results show that the azimuthal velocity profile is quite smooth \cite{bsj06}; no obvious Stewartson layer is observed at distances greater than $\sim1\;\unit{cm}$ from the bottom endcap \cite{se07}. The effect of the velocity jump across the junction between rings is not as severe as in the simulations reported here and those by \citet{hf04}. This difference may be explained by various instabilities associated with the Ekman and Stewartson layers. 

As for the Ekman layer, the low Reynolds number used in the simulations leads to a laminar Ekman-Hartmann layer as discussed in Sec.~\ref{sec:finite}. The experimental boundary-layer Reynolds number $Re_{\delta}=\sqrt{Re}\sim3\times10^{3}$ is larger than the critical value $Re_{crit}\sim10^{3}$ given $\Lambda\sim1$ \cite{gp71}, for the axisymmetric instabilities (viscous and inflection point instabilities) of Ekman-Hartmann layer with near-uniform rotation. And the vertical velocity shear due to the finite differential rotation in the experiment would result in the Kelvin-Helmholtz instability given a sufficiently high Reynolds number, which, however, could not be resolved by our axisymmetric simulations due to the same reason stated below. Therefore unstable Ekman layers are highly possible in the experiment. The layers may be smoothed by localized circulation and/or turbulence from these instabilities. 
This may account for the differences in the extent and prominence of the Stewartson layer between simulation and experiment.

As for the Stewartson layer, at the junction of the rings, the outer ring rotates more slowly than the inner one $(\Omega_{4}<\Omega_{3})$, hence $\partial(r^{2}\Omega^{2})/\partial r<0$ across the junction. This radial shear could result in both the Kelvin-Helmholtz instability and Rayleigh centrifugal instability given a sufficiently high Reynolds number. Unfortunately our axisymmetric simulation could not resolve the Kelvin-Helmholtz instability since it is a toroidal nonaxisymmetric mode. However, see \citet{fr99,hr03,hfme04,sc05} for experimental and theoretical studies of such instabilities in other contexts.

\begin{figure}
\begin{center}

  \scalebox{0.4}{\includegraphics{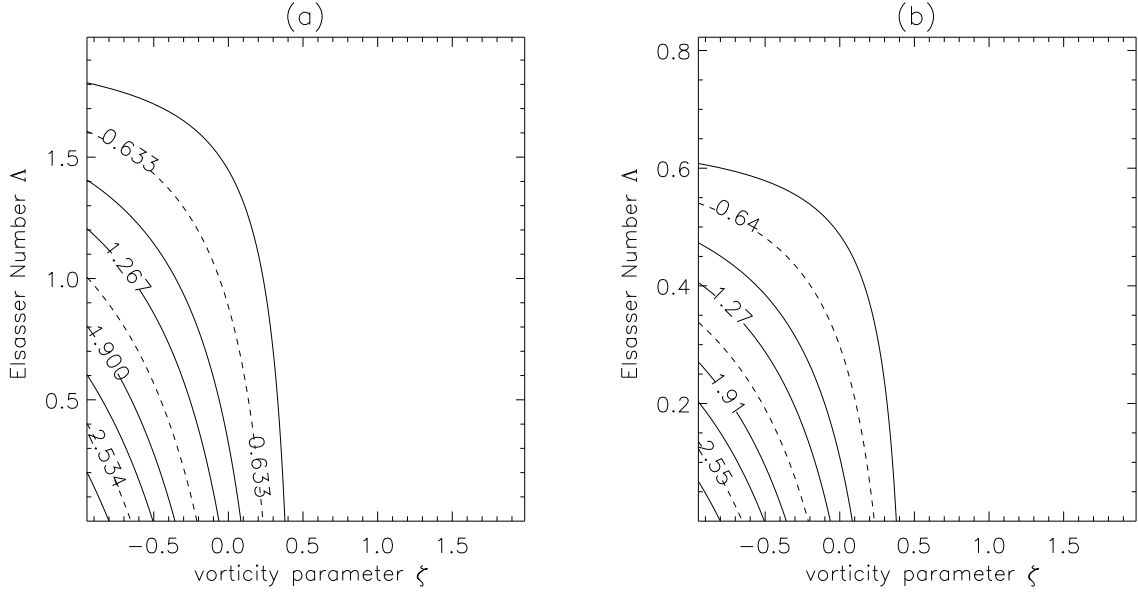}}
  \caption{\label{stewartson}  Growth rates ($\gamma\;\unit{s^{-1}}$) predicted by Eq.~\ref{disp} as a function of Elsasser number $\Lambda$ and vorticity parameter $\zeta$ at the junction of rings, $r_d=(r_1+r_2)/2$ with $Re=6400$. Parameters as in Table~\ref{Ta_parameter}. Panel (a) the vertical mode number $n=1$; panel (b) $n=2$.  }
\end{center}
\end{figure}

It is well known that surface tension at the interface between two fluids hinders the Kelvin-Helmholtz instability. In the real magnetized experiment, besides the instabilities discussed below, a less prominent Stewartson layer due to the Kelvin-Helmholtz instability would be resulted. In a homogeneous but magnetized fluid such as ours, magnetic field tension may stabilize the Kelvin-Helmholtz instability \cite{hs01,hr02}. However, the discussion of the nonaxisymmetric Kelvin-Helmholtz instability is beyond the scope of our current methods based on the axisymmetry and is not the main purpose of this paper.
Similarly magnetic field tension may stabilize the Rayleigh's centrifugal instability \cite{chan61}. 
We find that the short wavelength modes are stabilized before (\emph{i.e.}, at lower $\Lambda$) the long wavelength modes by performing a WKB stability analysis for the flows near the junction of the rings (Fig.~\ref{stewartson}), where the Stewartson layer lies. Our analysis assumes axisymmetry, thus the Kelvin-Helmholtz instability is excluded.
Following \citet{jgk01}, we have the dispersion relation:
\begin{equation}
\label{disp}
[(\gamma+\nu k^2)(\gamma+\eta k^2)+(k_{z}V_{A})^2]^2\frac{k^2}{k_{z}^{2}}+\kappa^2(\gamma+\eta k^2)^2+\frac{\partial \Omega^2}{\partial \ln r}(k_{z}V_A)^2=0
\end{equation}
All variables have the same meanings as in \citet{jgk01} except: (1) the characteristic rotation speed $\Omega$ is chosen to be $\sqrt{\Omega_{3}\Omega_{4}}$; (2) the dimensionless vorticity parameter, $\zeta\equiv(1/r\Omega)\partial(r^2\Omega)/\partial r=2+\partial \ln \Omega/\partial \ln r$ is taken to be $2+(r_d/\Omega)(\Delta\Omega/\Delta r)$, where $\Delta\Omega=\Omega_3-\Omega_4$ and $\Delta r$ is the radial thickness of the Stewartson layer; (3) the wave number $k=\sqrt{k_{z}^2+k_{r}^{2}}$, where the axial wave number $k_{z}=n\pi/(h/2)$ and $k_{r}=\pi/\Delta r$ ($n$ is the vertical mode number) since the radial and axial characteristic lengths are $\Delta r$ and $h/2$, respectively. 

From Fig.~\ref{stewartson}, the Rayleigh's centrifugal instability, which occurs for all $\zeta<0$ when $V_{A}=0$ and $\nu=0$, is found to be suppressed by a strong magnetic field. This is consistent with \citet{chan61}. It is very interesting to see some growing modes when $\zeta>0$, which is the MRI associated with the Stewartson layer. This instability also disappears with a sufficiently strong magnetic field. Comparing panel (a) with panel (b), we find that the magnetic field suppresses the modes with shorter wavelengths more strongly. This could explain why the Stewartson layer extends deeper into the bulk with a stronger magnetic field (Fig.~\ref{ring2_ekman_vt}), by suppressing the growing modes with short wavelengths that would otherwise tend to smooth the velocity gradient. 

\begin{figure}
\begin{center}

  \scalebox{0.4}{\includegraphics{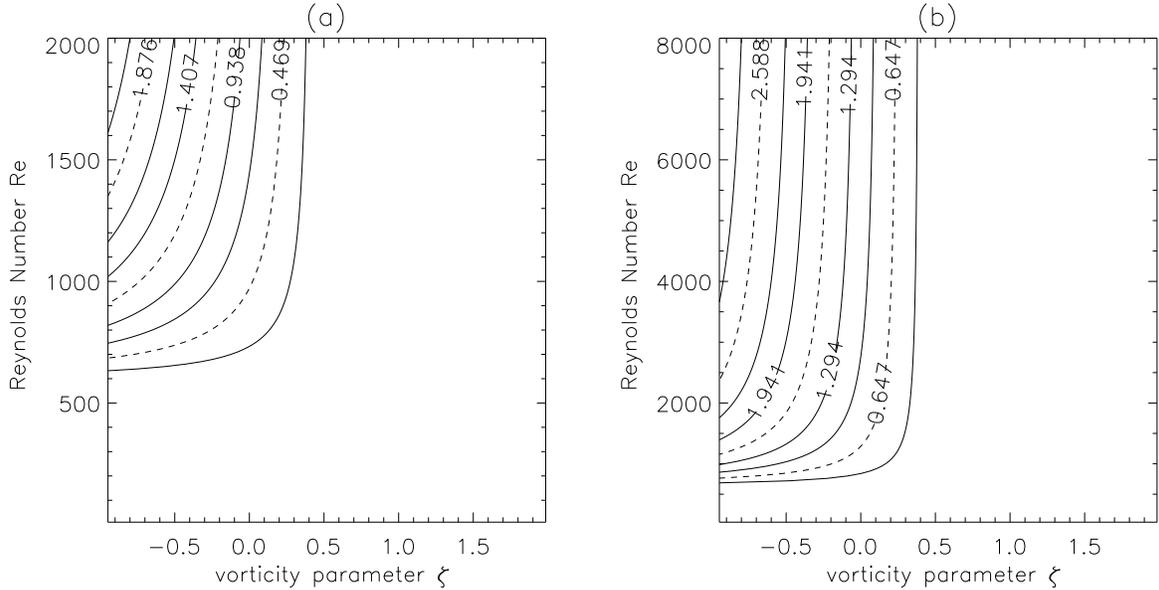}}
  \caption{\label{stewartson2}  MRI linear growth rates ($\gamma\;\unit{s^{-1}}$) as a function of Reynolds number $Re$ and vorticity parameter $\zeta$ at the junction between rings, $r_d=(r_1+r_2)/2$ with $\Lambda=1.5$. Parameters as in Table~\ref{Ta_parameter}. Panel (a) the vertical mode number $n=0.25$; panel (b) $n=1$. }
\end{center}
\end{figure}

Viscosity also has a stabilizing influence (Fig.~\ref{stewartson2}). Given $\Lambda=1.5$, the Stewartson layer is found to be stable if $Re\lesssim600$ regardless of the vertical mode number. This could explain why the Stewartson layer penetrates deeper into the bulk with increasing Reynolds number if the layer is steady (Fig.~\ref{zeta} (b), $Re\lesssim400$). However at even larger Reynolds number the instabilities would destabilize the layer and presumably smooth it out except near the endcaps (Fig.~\ref{zeta} (b), $Re\gtrsim400$). If we could perform simulations without magnetic fields at the experimental Reynolds number ($Re\gtrsim10^7$), we expect that the profile of the azimuthal velocity \emph{v.s.} radius at $\sim1\;\unit{cm}$ above the bottom endcap would be a tiny hump with a large slope of the azimuthal velocity $|\partial v_{\varphi}/\partial r|$ so that our experimental measurement would not resolve it, which matches the experimental evidence that an obvious Stewartson layer is not observed in a purely hydrodynamic experiment \cite{bsj06}. Unfortunately the current ZEUS code cannot afford a simulation with Reynolds number as high as the one in the experiment. 

\acknowledgments
The authors would like to sincerely thank Jeremy Goodman and Hantao Ji for their very inspiring discussion and constructive comments.  The authors would also like to thank James Stone for the advice on the ZEUS
code.This work was supported by the US Department of Energy, NASA under grants ATP03-0084-0106 and APRA04-0000-0152 and also by the
National Science Foundation under grant AST-0205903.

\end{document}